\documentclass[twocolumn,aps,showpacs,prl,epsf,graphics,psfig]{revtex4}
\usepackage{graphicx}

\begin{document}
\title{
Optical control of DNA-base radio-sensitivity
}

\author{Ramin M. Abolfath}

\affiliation{
Department of Radiation Oncology, University of Texas,
Southwestern Medical Center, Dallas, TX 75390
}

\date{\today}

\begin{abstract}
{\bf Purpose}: Manipulation of the radio-sensitivity of the nucleotide-base driven 
by the spin blockade mechanism of diffusive free radicals against ionizing 
radiation. \\
{\bf Materials and methods}: We theoretically propose a mechanism which uses the 
simultaneous application of circularly polarized light and an external 
magnetic field to control the polarization of the free radicals and create 
S=1 electron-hole spin excitations (excitons) on nucleotide-base. We deploy 
an ab-initio molecular dynamics model to calculate the characteristic 
parameters of the light needed for optical transitions. \\
{\bf Results}: As a specific example, we present the numerical results calculated 
for a Guanine, in the presence of an OH free radical. To increase the 
radio-resistivity of this system, a blue light source for 
the optical pumping and induction of excitons on guanine 
can be used. \\
{\bf Conclusions}: The effect of spin-injection on the formation of a free energy 
barrier in diffusion controlled chemical reaction pathways leads to the 
control of radiation-induced base damage. The proposed method allows us to 
manipulate and partially suppress the damage induced by ionizing radiation.
\end{abstract}
\maketitle

\section{Introduction}
Ionizing radiation is both hazardous and beneficial to living organisms, 
and is extensively used for cancer treatment in radiation 
therapy~\cite{Khan:book}. 
A major problem in the application of ionizing radiation to cancer treatment 
is the protection of normal cells and tissues against unavoidable exposure to 
radiation during radiation treatment. 
It is now well understood that the ionization or excitation of the DNA 
molecules, either directly or indirectly, can lead to DNA single or double 
strand breaks.
As a result, misrepaired DNA molecules can lead to specific genetic aberrations and/or mutations 
which could cause carcinogenesis in normal cells or lead to fatal damage in 
normal or cancer cells ~\cite{Lyudmila2008:Science,EricHall:book}.
It has been shown that low linear-energy-transfer (LET) ionizing radiation 
creates approximately 1,000 
single strand breaks (SSBs) and 40 double strand breaks (DSBs) per gray 
(1Gy=1J/Kg) in typical mammalian 
cells~\cite{Ward1988:RR,Goodhead1994:IJRB,Nikjoo1997:IJRB,Semenenko2004:RR}.
The level of DNA molecular base damage 
is around 2,500 to 25,000 per Gy in a cell,
which is about 2.5 to 25 times the yield of sugar-phosphate induced damage 
in the DNA backbone~\cite{Ward1988:RR,Semenenko2004:RR}.
In indirect mechanisms, the water molecules surrounding the DNA molecule 
which compose 80\% of a cell, 
may be excited by ionizing radiation in form of free radicals, 
e.g., a charged neutral hydroxyl (OH).
The motion of OH-radicals which are randomly produced throughout the cell 
is governed by diffusion processes. 
Massive DNA damage can result from a large number of
DNA dehydrogenations caused by free radicals. 
For example, a free radical can diffuse to reach a DNA molecule and
remove a hydrogen ion from it to form a water molecule. 
Detailed studies at the molecular level is necessary to bring
the radiation-induced DNA damage under control.
%

\section{Method}
In this work, we apply a quantum physical description of molecular 
interactions to propose a mechanism that could allow the manipulation 
of DNA radio-sensitivity. 
In particular the Pauli exclusion principle~\cite{LandauLifshitzQM:book} 
which  prevents two electrons with parallel spin form occupying a single 
spatial orbital, plays a major role and is used to magnetically manipulate 
the diffusion of hydroxyl radicals and the OH-DNA relative motion.
It has been shown in studies in semiconductor physics and quantum optics 
that the Pauli exclusion principle can be used to rectify electrical currents 
passing through weakly coupled quantum dots ~\cite{Ono2002:Science} and
to induce ferromagnetic ordering by photo-generated carriers in magnetic 
semiconductor hetero-structures~\cite{Oiwa2002:PRL}. 

A free radical carries an odd number of electrons with an unpaired spin in 
the outermost open shell.
Due to the reduction of the exchange interaction, the pairing of opposite
spin electrons in the open orbital of the free radical with an electron in 
a DNA molecule makes free radicals highly reactive.
In the process of dehydrogenation of a DNA molecule by free radicals 
an unpaired hole (a half-empty orbital) is transferred to the DNA.
In the absence of spin-orbit coupling and hyperfine interaction   
the spin of transfered electron is conserved.
The electronic ground state of DNA-molecule is $S=0$
spin-singlet (in the absence of an external magnetic field).
The OH-radical which contains nine electrons is a doubly  degenerate  
ground state with $S_z=\pm 1/2$, 
where we have conveniently taken the quantization axis along the $z$-axis.
The degeneracy of the ground state can be lifted by 
applying a weak magnetic field
which couples to the electron spin through the  
Zeeman interaction \cite{LandauLifshitzQM:book}, 
$E_{Z}= g\mu_B \vec{S}\cdot \vec{B}$.
Here $E_{Z}$ is the Zeeman energy, 
$g$ is the electron $g$-factor ($g\approx 2$), 
$\mu_B$ is the Bohr magneton ($\mu_B\approx 5.8\times 10^{-5}$ eV/Tesla), 
and $B$ is the strength of external magnetic field.

In a random interaction of radiation with a biological system 
the initial direction of OH-radical magnetic moment 
immediately after its generation is also random.
However, by applying a weak external magnetic field ($B_{\rm ext}$) 
(which defines the quantization axis)
and using a circularly polarized light field 
parallel to the direction of the light propagation, as shown in Fig.\ref{Fig0}a,  
a molecular transition corresponding to $\Delta J=\pm 1$
can be induced by means of optical pumping \cite{Walker1997:RMP} of  the 
OH-radicals~\cite{Meerakker2005:PRL}.
Here $J$ denotes the total angular momentum of diatomic
OH-radical \cite{Herzberg:Bppk}.
Alternatively, tchniques such as electron spin resonance (ESR) can be used to 
achieve strong polarization of free radicals, as recent advances in ESR have
demonstrated the capability of detecting the transfer of electron spin 
polarization between radicals \cite{Jenks1999:JACS,Brocklehurst1979:FT}.
In this case microwaves can be used for the optical transitions.
In a similar fashion, by applying a second circularly polarized light field 
one may excite an electron-hole pair (exciton) in the DNA molecule.
Because the circularly polarized light carries angular momentum $\pm 1$, 
the exciton has a particular spin polarization.
Here the spin of exciton is $S=1$ with polarization along 
the light propagation direction (because of angular momentum 
selection rules).
Fig. \ref{Fig0} schematically shows the generation of the
optically pumped exciton by circularly polarized light.
The injection of photo-electrons with the spin out of equilibrium may lead
to a dramatic effect in the collective dynamical behavior of  
DNA-molecules and the interaction with OH-radicals.
For example the OH-DNA repulsive magnetic force provides a potential
barrier which blocks the diffusion pathway (see Fig. \ref{Fig0}b)
of OH-radicals toward the DNA-molecules. 
This is expected to hinder the DNA dehydrogenation  
and consequently increase the cell radio-resistivity. 
To verify this hypothesis, an {\em ab}-initio molecular dynamical model, 
which is the mathematical formulation that governs the appropriate 
dynamics of the molecular system~\cite{Marx_Hutter} is deployed.
We have used the Car-Parrinello molecular dynamics 
(CPMD)~\cite{Car1985:PRL,CPMD} model, 
in which the potential energy of the system can be calculated on-the-fly, 
as needed for the conformations of the dynamical trajectory 
to simulate the chemical reaction pathways.
Because the absorption of a circularly polarized photon alters the 
local electronic state of a DNA-molecule,
we confine our simulation to a particular segment, e.g., only a part of the 
DNA where the injected exciton 
is localized and the optical transition takes place.
To illustrate this, let us consider a system of interest consisting of 
a DNA nucleotide base, (e.g., guanine) in the presence of the OH-radical. 
We assume that a photon with circular polarization interacting with guanine 
can induce an optical transition in the form of an $S=1$ exciton.
Here we investigate the effect of an exciton produced in this way 
on the guanine-dehydrogenation pathway, assuming that another photon 
generated through interactions with ionizing radiation 
(such as radiotherapy x-rays or cosmic rays) 
creates a free radical in the vicinity of guanine.
Because the local density of free radicals and excitons are large and are
comparable, the events described in our calculation can be observed with 
reasonable probability.
We adopt computational parameters and variables needed for the CPMD calculation 
of the dynamical trajectory of the gas phase nucleotide bases 
in the presence of OH-radicals
following Refs. \cite{Mundy2002:JPC}, where 
the consistency of CPMD results for guanine 
with other quantum chemistry approaches has been investigated.

\section{Results}
\begin{figure}
\begin{center}
\includegraphics[width=0.8\linewidth]{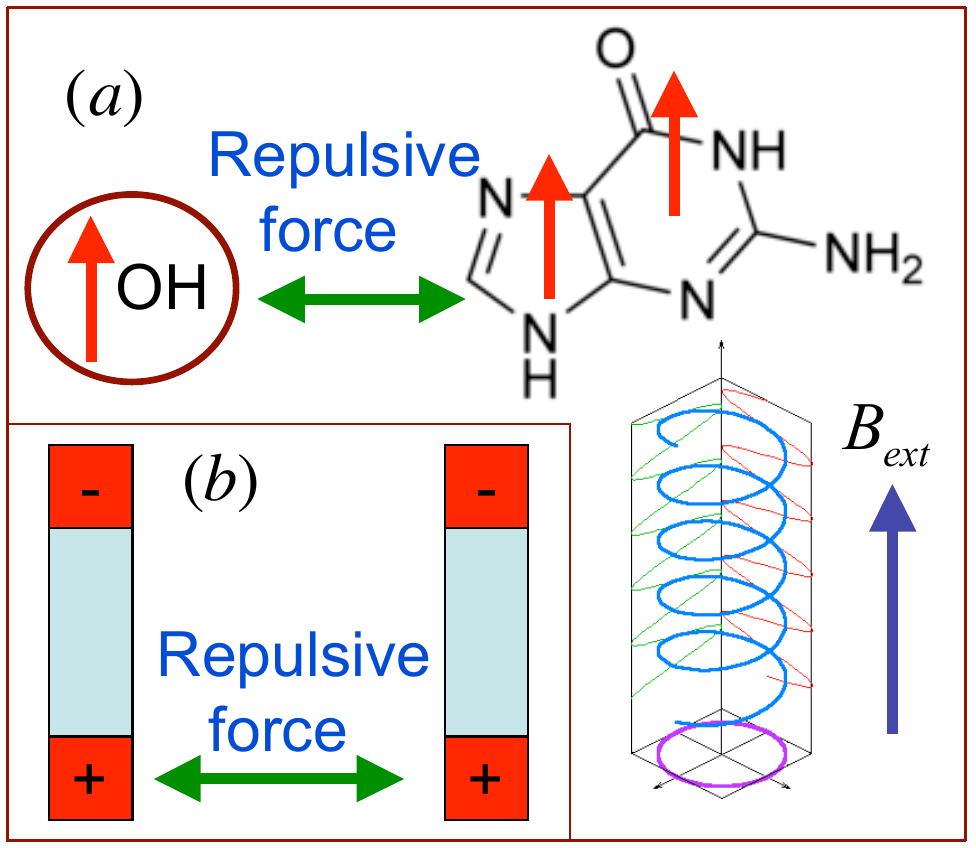}
\noindent
\caption{
Schematically shown in (a) the injection of photo-generated electrons in 
DNA-nucleotides with spin polarization (shown by arrows) along the direction 
of circularly polarized light and external magnetic field. The net magnetic 
force between two parallel magnetic moments localized in OH and DNA-nucleotide 
is repulsive. This is similar to two separated magnetic moments which interact 
like Heisenberg antiferromagnetic exchange coupling (b).
}
\label{Fig0}
\end{center}
\end{figure}

We identify the dehydrogenation of the nucleotide bases 
as a function of their spin multiplicity.
The ground and excited states of the nucleotide correspond to 
spin singlet ($S=0$), and spin triplet ($S=1$) states.
The latter can be realized through the application of circularly polarized 
light as discussed above (see Fig. \ref{Fig0}). 
Our CPMD is implemented in a plane-wave basis within local spin density
approximation (LSDA) with an energy cutoff of 70 Rydberg (Ry), and 
with Becke \cite{Becke1988:PRA} exchange and Lee-Yang-Parr (BLYP) 
gradient-corrected functional \cite{Lee1988:PRB}.
Norm conserving ultrasoft Vanderbilt pseudo-potentials were used for 
oxygen, hydrogen, nitrogen and carbon.
The CPMD micro-canonical dynamics (constant energy ensemble) 
were performed after wave-function optimization following dynamical 
equilibration at T=300K and re-quenching of the wave-function.
An isolated cubic cell of length 13.229 $A$ with Poisson solver of Martyna and
Tuckerman \cite{Martyna1999:JCP} was used. 
Our CPMD studies consist of two classes of spin-restricted calculations, as
the total spin along the quantum axis 
is subjected to the constraints $S_z=1/2$, and $3/2$,
corresponding to doublet and quartet spin configurations.
In both calculations the initial distance between OH-radical and nucleotide 
is considered to be about 1.5 $A$. 
We selectively choose an initial coordinate for OH-radical in the 
neighborhood of the nucleotide
where the Hydrogen transfer shows a reactive path 
in normal state of DNA (the doublet spin configuration in the absence of
circularly polarized light and magnetic field).

\begin{figure}
\begin{center}
\includegraphics[width=0.8\linewidth]{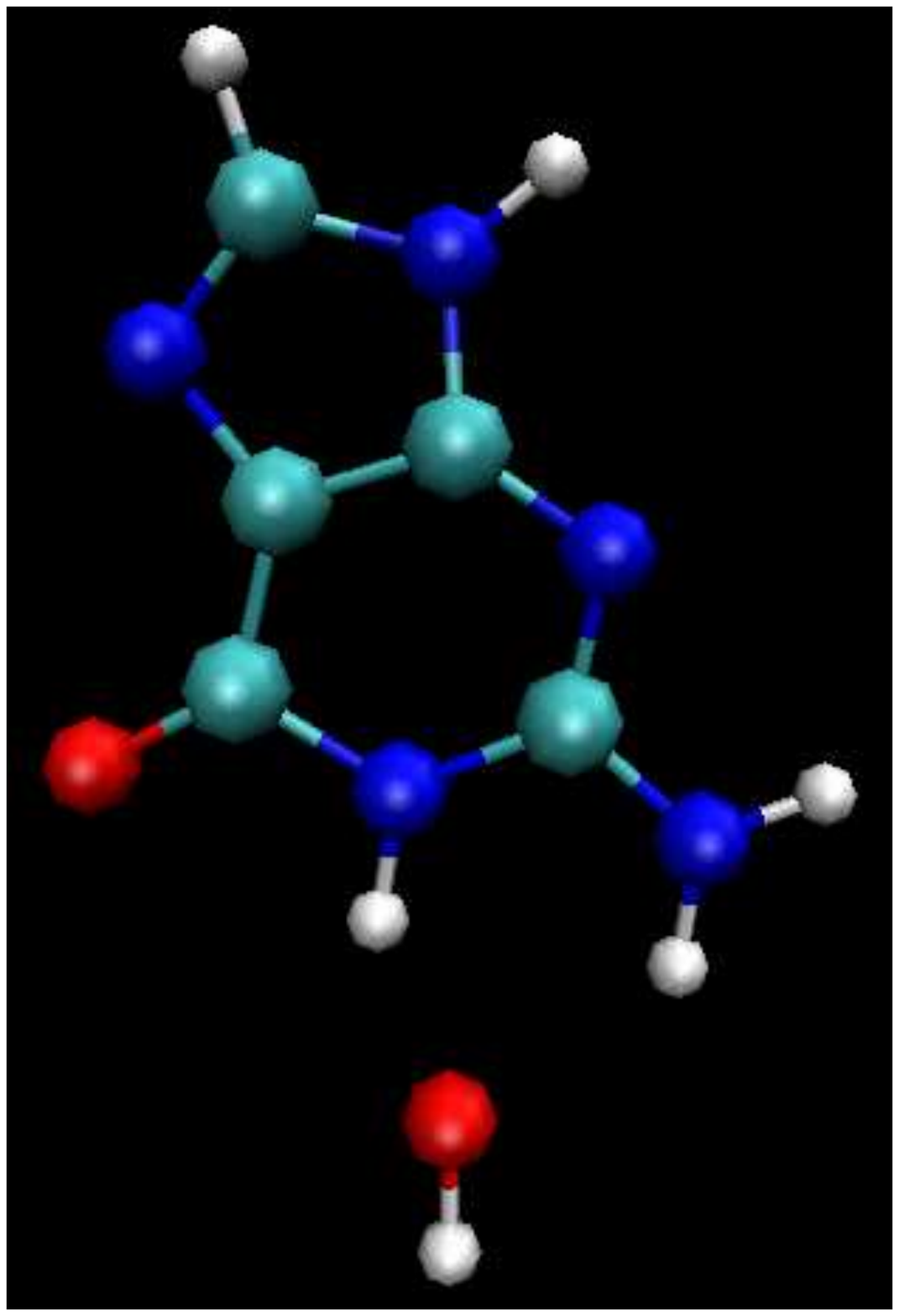}
\noindent
\caption{
Initial state of Guanine molecule in the presence of irradiated induced
OH free radical.
}
\label{Fig1a}
\end{center}
\end{figure}

\begin{figure}
\begin{center}
\includegraphics[width=0.8\linewidth]{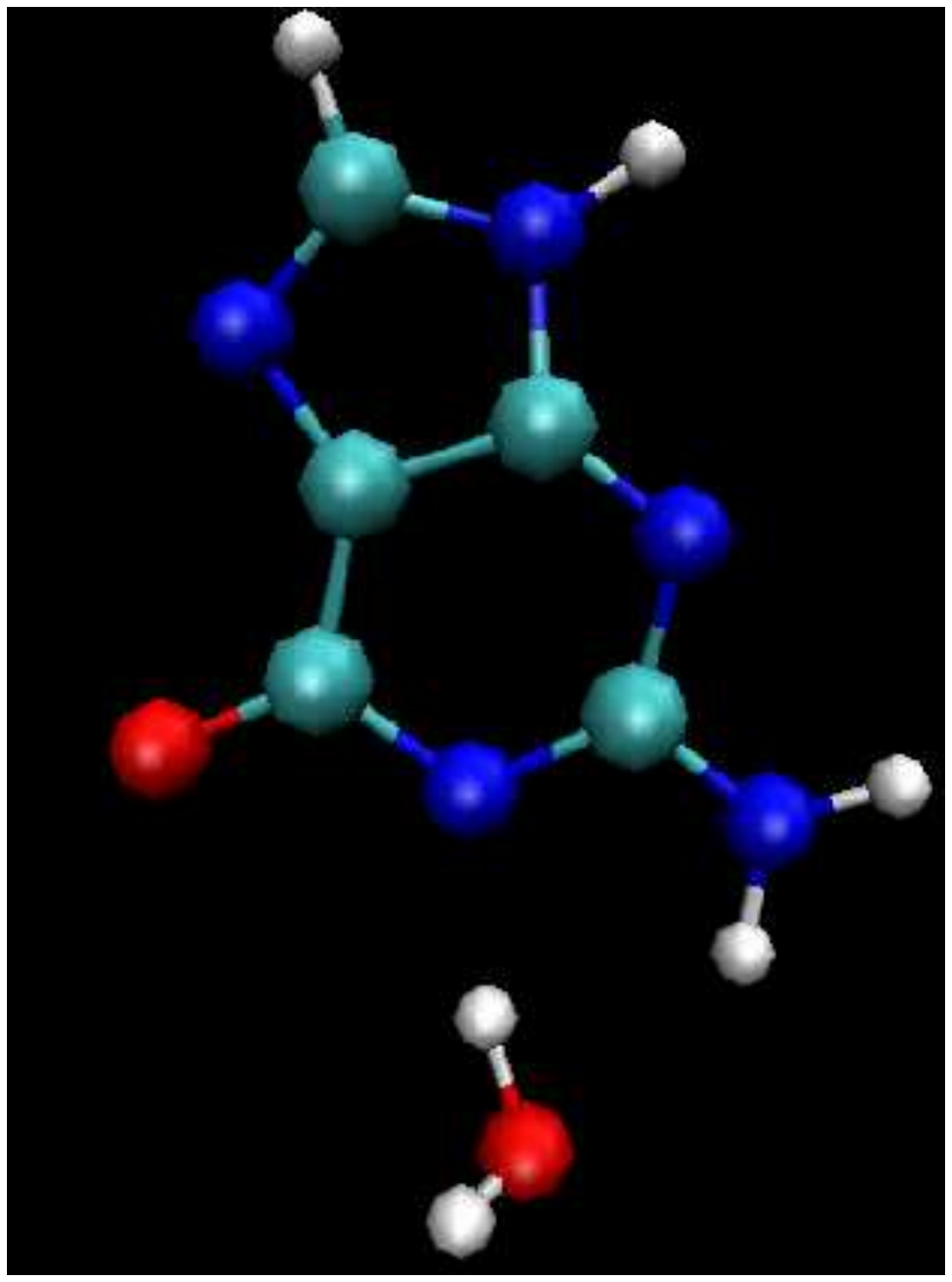}
\noindent
\caption{
The state of de-hydrogenated Guanine by OH free radical at $t=0.6$ ps.
The polarization state of the system is spin doublet ($S=1/2$).
}
\label{Fig1b}
\end{center}
\end{figure}

\begin{figure}
\begin{center}
\includegraphics[width=0.8\linewidth]{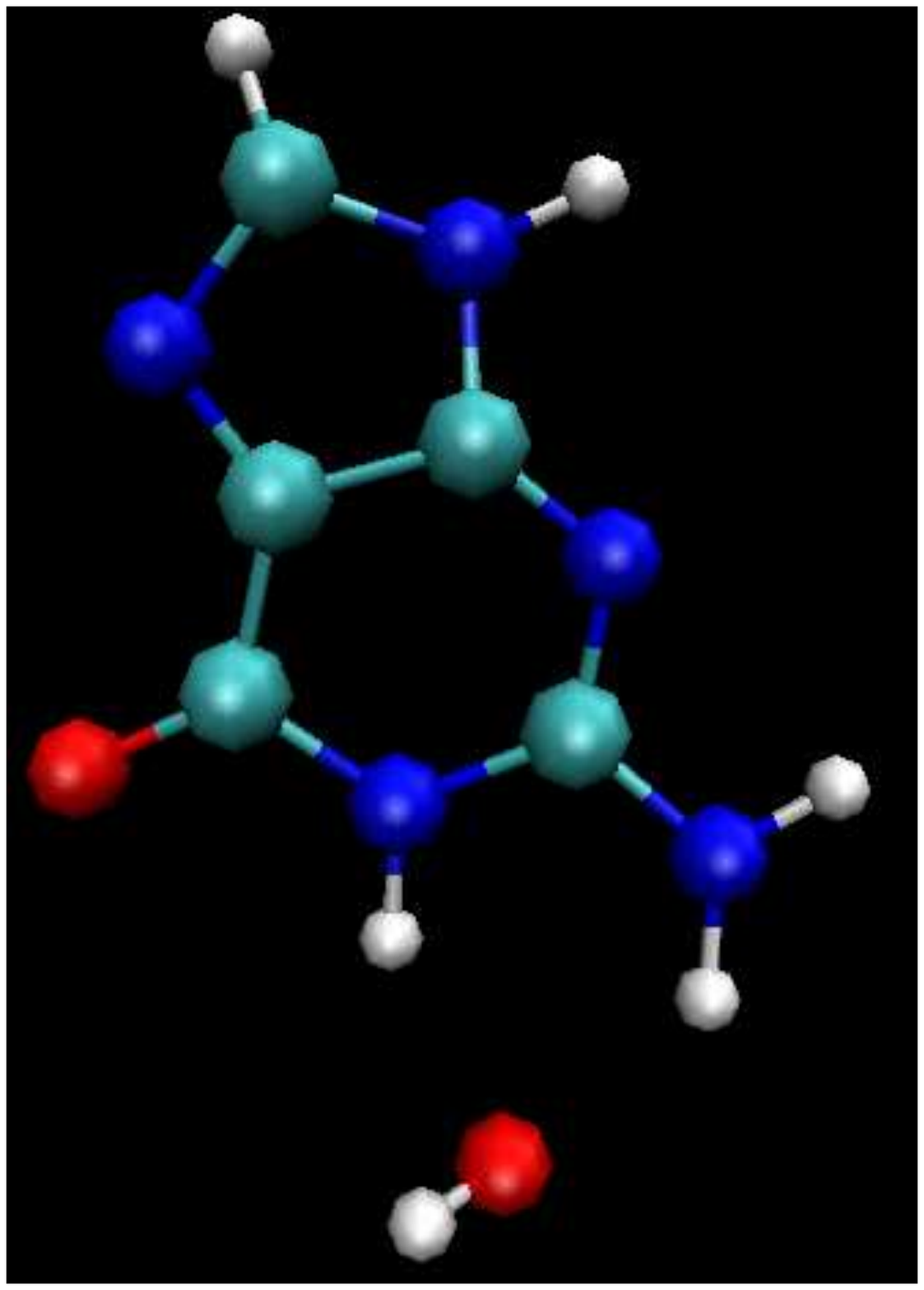}
\noindent
\caption{
The state of radio-resistive Guanine at $t=0.6$ ps.
The polarization state of the system is spin quartet ($S=3/2$) induced by
circularly polarized light in the presence of weak magnetic field.
Due to injected polarized photo-electrons localized in Guanine, the 
dehydrogenated Guanine does not form.
}
\label{Fig1c}
\end{center}
\end{figure}

\begin{figure}
\begin{center}
\includegraphics[width=1.0\linewidth]{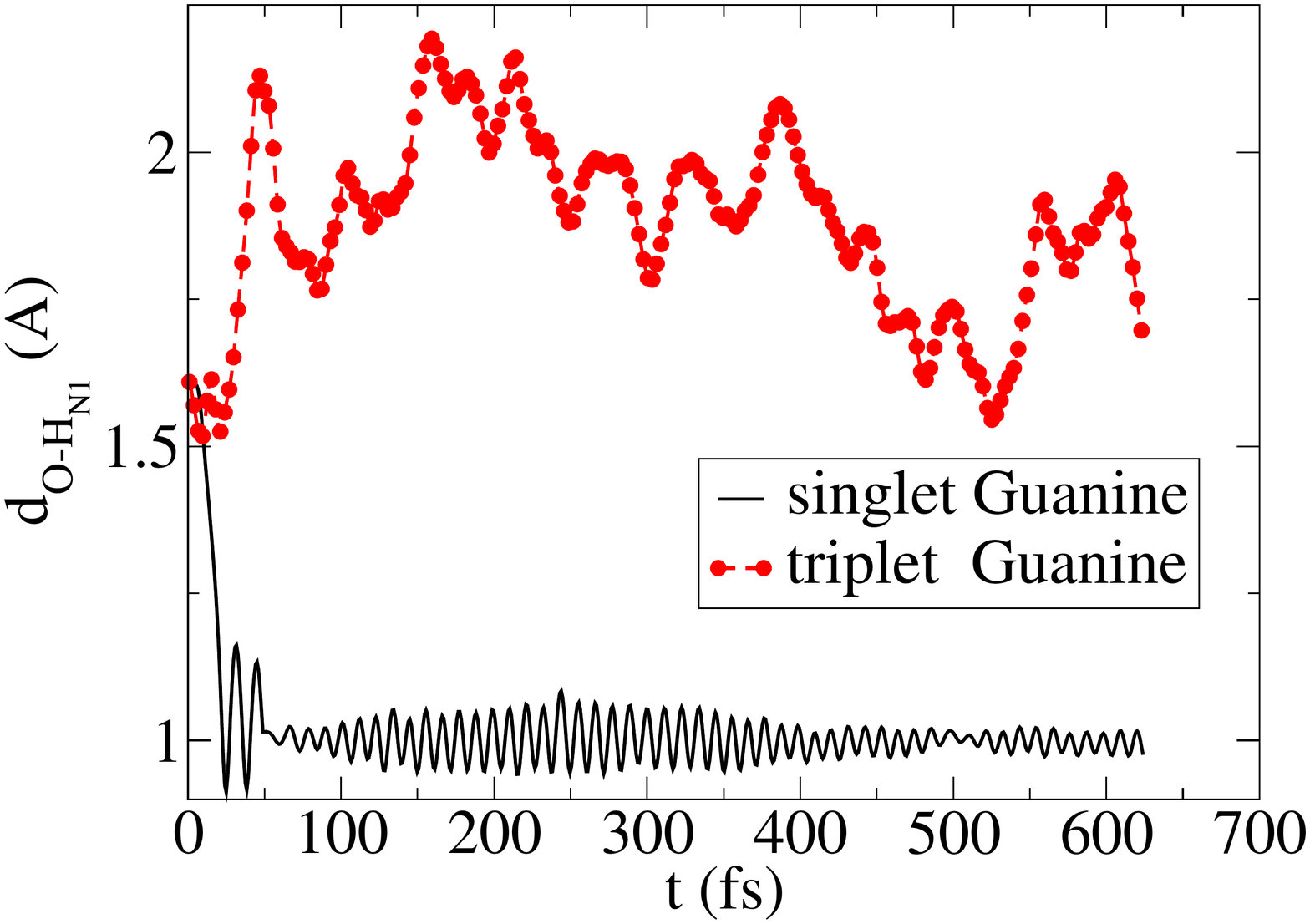}
\noindent
\caption{
The evolution of the distances in Angstrom from oxygen atom in the OH
radical to the H$_{N1}$ in the guanine as a function of guanine spin
multiplicity. 
The hydrogen abstraction occurs around $t\approx 50$fs in 
spin singlet state of guanine.
The injection of $S=1$ exciton in guanine blocks the hydrogen abstraction.
}
\label{Fig5}
\end{center}
\end{figure}

\begin{figure}
\begin{center}
\includegraphics[width=1.0\linewidth]{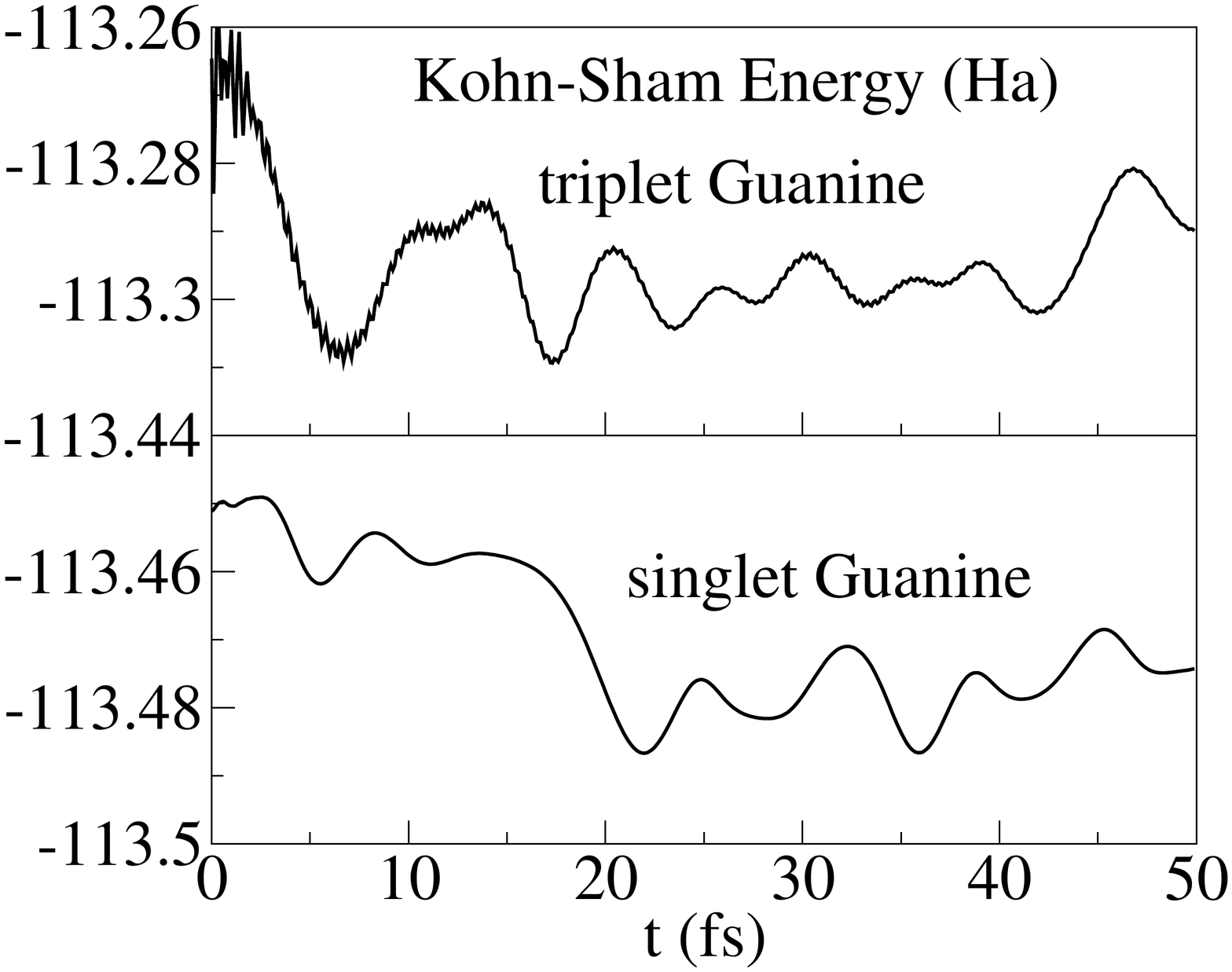}
\noindent
\caption{
The Kohn-Sham energy as a function of time and spin multiplicity 
of Guanine, spin triplet (top) and spin singlet (bottom).
A drop in Kohn-Sham energy in spin singlet multiplicity is indication
of dehydrogenation of H$_{N1}$ in the guanine by OH free radical.
}
\label{Fig6}
\end{center}
\end{figure}

The initial and final states of the molecules are shown in
Figs. \ref{Fig1a}-\ref{Fig1c}. 
The final configurations of the molecules have been obtained after 0.6 ps
where the rearrangement of the atomic coordinates have been
deduced from a dynamical trajectory calculated by CPMD.
According to our results, a rapid dehydrogenation 
of the nucleotides takes place 
for a system with $S_z=1/2$ (total spin-doublet) as shown in Fig. \ref{Fig1b}.
This process leads to the formation of a water molecule.
In contrast, as shown in Fig. \ref{Fig1c}, 
in the quartet spin configuration the repulsive exchange 
interaction, analogous to Heisenberg anti-ferromagnetic coupling which 
originates 
from the Pauli exclusion principle, blocks the exchange of hydrogen
and hence the chemical reaction.
In Fig. \ref{Fig5} the evolution of the N$_1$ Hydrogen in the guanine 
and free radical oxygen distance is shown.
As it is seen the abstraction of Hydrogen occurs around $t\approx 50$fs in 
the spin singlet state of guanine, and
the injection of $S=1$ exciton in guanine blocks the hydrogen abstraction.
Fig. \ref{Fig6} shows the Kohn-Sham energies (equivalent to potential
energy in classical molecular dynamics) of the spin singlet
and spin triplet of the Guanine in the presence of the OH free radical
as a function of time, calculated by
the CPMD at T=300K corresponding to a canonical dynamics 
(constant temperature ensemble).
A drop in Kohn-Sham energy in spin singlet multiplicity is indication
of dehydrogenation of H$_{N1}$ in the guanine by OH free radical.
To systematically check the convergence of the results, we 
increased the size of the molecule by adding sugar-phosphate  
rings to guanine and found that this has no influence on the spin-blocking 
effect.
To estimate the energy needed for the polarization of the nucleotide
in the absence of OH-radicals,
we calculated the energy of the ground and excited states of the gas-phase 
nucleotide in spin singlet and triplet multiplicities.
For guanine we calculated the spin singlet-spin triplet energy gap 
$\Delta_0 \equiv E_{\rm triplet} - E_{\rm singlet} \approx 2.68$ eV. 
This provides an estimate for the frequency of the 
circularly polarized light, which is within the range of the  
visible spectrum of the electromagnetic waves, $\lambda=463$ nm (light blue). 
To calculate the stored magnetic energy due to the optical injection of spin,
we calculated the energy of the gas-phase nucleotide 
in the presence of one OH-free radical with spin doublet and 
quartet multiplicities.
For the molecules shown in Fig. \ref{Fig1a}, we find the energy gap 
$\Delta_1 \equiv E_{\rm quartet} - E_{\rm doublet} \approx 3.54$ eV.
Here the excessive magnetic energy which originated from 
spin-spin repulsive interactions (which resemble the anti-ferromagnetic 
exchange interaction in the Heisenberg model) can be deduced to be 
$\Delta_1 - \Delta_0 \approx 0.86$ eV.
This energy can be interpreted as the excessive energy barrier
due to the alignment of the spins in the DNA molecule and OH, and is
the source of the magnetic repulsive force which makes the diffusion of OH  
toward DNA-molecules less likely. 
This is in agreement with the results obtained from CPMD, shown in
Figs. \ref{Fig1a}-\ref{Fig1c}. 
In addition, by switching the polarization of one of the light sources
to the opposite direction,  
the relative direction of the DNA-OH polarization switches to 
antiparrallel, and hence the
magnetic repulsive force changes to an attractive  
force that lowers the OH diffusion barrier and 
decreases the radio-sensitivity of the DNA-molecule.

After photon absorption, the nucleotide is spin polarized along
the direction determined by the polarization state and the propagation
direction of the circularly polarized light.
The polarized state of the nucleotide then decays quantum mechanically
to its unpolarized ground state either by spontaneous photon emission 
(electron-hole recombination) or through photon-electron spin decoherence.
There are radiative and non-radiative channels that contribute 
to this process. 
Since spin-orbit coupling governs one of the decay mechanisms in non-radiative 
channel, we use Fermi's golden rule to estimate the life-time of the triplet 
state.
It then follows that $\Gamma_{T\rightarrow S}=(2\pi/\hbar^2) \Omega
\int d^3k/(2\pi)^3 
|\langle T| H_{SO}|S\rangle|^2 \delta(\omega_0 - \omega)$
\cite{LandauLifshitzQM:book}. 
Here, $\Gamma_{T\rightarrow S}$ is the transition rate from the spin-triplet 
($T$) to the spin-singlet ($S$) state, $\Omega$ is the volume, 
$k=2\pi/\lambda$ is the emitted photon wave-number,
$H_{SO}=-e/(2m^2 c^2) \sum_{i=1}^N \vec{s}_i\cdot(\vec{p}_i
\times\vec{\nabla} \Phi_{KS}(\vec{r}_i))$ is the spin-orbit Hamiltonian,
$m$ is the electron mass, $c$ is the speed of light, $s_i$, $p_i$ are the spin
and momentum of the $i$th electron, $\Phi_{KS}$ is the
Kohn-Sham effective potential, and $\omega_0 = \Delta_0/\hbar$.
This calculation shows that $\tau=\Gamma^{-1}_{T\rightarrow S} \approx 100$ ps.
However, the electronic relaxed excitonic states with emperical lifetime of 
several 100 ps has been reported recently~\cite{Buchvarov2007:PNAS}.
The spin-triplet life time of nucleotide $\tau$ turns out to be significantly
larger than the dehydrogenation time scale.
It is therefore possible to increase the radio-resistivity of the
DNA molecule within this time scale through optically pumped spin polarization.
It is important to compare $\tau$ with other
time-scales in the process. 
The initial ionization takes place in about 1 fs ($10^{-15}$ second).
The primary free radicals produced by ejection of electrons have a life time of
nearly 100 ps, and the reported OH radical life-time is about 1 ns 
($10^{-9}$ second)~\cite{EricHall:book}.
The  electron spin-lattice relaxation time of the OH radical has 
been estimated to be approximately between 0.1 and 0.5 ns in water at room 
temperature~\cite{Brocklehurst1979:FT}. 

In order to estimate the technical requirements for the above described 
approach one could assume that an aqueous solution of DNA will be irradiated 
with a dose of 1Gy (1J/Kg). 
It is known~\cite{Hiroshi2005:JRR} that 100eV of absorbed photon/electron 
energy produce about 6 OH radicals. 
Therefore 1Gy of radiation produce about $4 \times 10^{13}$ OH radicals in 
0.1 cm$^3$ of water.  
If the number of injected excitons can be exceeded up to at least ten 
times, by applying a laser pump with moderate intensity 
it is possible to increase significantly the resistance of DNA-molecules 
against irradiation.
For example at a dose rate of 1.4 Gy/min a laser pump
power of $P=10 N_{OH} \hbar\omega_0 /\tau_{ir} \approx 12 \times 10^{-6}$ 
watt would be required, 
which is well within the technically achievable limits.

\section{Conclusion}
In conclusion,  we have theoretically explored a mechanism 
which involves the injection of spin polarized excitons
in DNA molecules to control and manipulate the
radio-sensitivity of cells by using a 
circularly polarized light field and external magnetic field.
The mechanism proposed here is based on
the selection rules applicable to optical transitions between 
energy levels of the DNA-molecules and optical pumping
of the OH-radicals, and we have employed a microscopic 
{\em ab}-initio molecular dynamics model to computationally 
study the dehydrogenation mechanism at the molecular level.
The results of this study may be used as a guideline to develop new 
techniques for  radiation therapy and radiation protection purposes. 

The author thanks Homayoun Hamidian, Reinhard Kodym, Lech Papiez, 
and Tim Solberg for the comments and useful discussions.


       


\end{document}